\documentclass{article}

\usepackage{graphicx}
\usepackage[left=3cm,right=2cm,top=2cm,bottom=2cm,nohead,nofoot]{geometry}
\usepackage[english]{babel}
\usepackage{amsmath}
\usepackage{amssymb}
\usepackage{color}
\usepackage{hyperref}
\usepackage{cite}
\usepackage{setspace}
\newcommand{\rmd}{\mathrm{d}}

\newcommand\figref[1]{Fig.~\ref{#1}}

\newcommand{\subfigf}{0.4\textwidth}
\newcommand{\subfigff}{0.3\textwidth}
\newcommand{\subfigfff}{0.3\textwidth}

\title{Burst and inter-burst duration statistics as empirical test of long-range memory in the financial markets}

\author{V.~Gontis, A.~Kononovicius}

\date{\small Institute of Theoretical Physics and Astronomy, Vilnius University\\Saul{\. e}tekio al. 3, 10257 Vilnius, Lithuania}

\begin{document}

\maketitle

\begin{abstract}
We address the problem of long-range memory in the financial markets. There are two conceptually different ways to reproduce power-law decay of auto-correlation function: using fractional Brownian motion as well as non-linear stochastic differential equations. In this contribution we address this problem by analyzing empirical return and trading activity time series from the Forex. From the empirical time series we obtain probability density functions of burst and inter-burst duration. Our analysis reveals that the power-law exponents of the obtained probability density functions are close to $3/2$, which is a characteristic feature of the one-dimensional stochastic processes. This is in a good agreement with earlier proposed model of absolute return based on the non-linear  stochastic differential equations derived from the agent-based herding model.
\end{abstract}

\section{Introduction}
The long-range memory in volatility and trading activity of the financial markets is still one of the most mysterious features attracting permanent attention of researchers \cite{Engle2001QF,Plerou2001QF,Gabaix2003Nature,Ding2003Springer} as some ambiguity is still present.
There are numerous works, which construct autoregressive models with embedded long-range memory -- such as FIGARCH, LM-ARCH, FIEGARCH  \cite{Giraitis2007,Giraitis2009,Conrad2010,Arouri2012,Tayefi2012}. 
The slowly decaying auto-correlation for volatility and trading activity, is a characteristic property of the empirical financial time series \cite{Plerou2001QF,Gabaix2003Nature,Ding2003Springer}. This empirical property might originate from the true long-range memory process with correlated increments such as the fractional Brownian motion (fBm) \cite{Mandelbrot1968SIAMR,McCauley2006PhysA,McCauley2007PhysA} or from the Markov processes with slowly decaying auto-correlation such as stochastic processes with non-stationary uncorrelated increments \cite{Gontis2004PhysA,McCauley2006PhysA,McCauley2007PhysA,Ruseckas2011PRE}. The debate whether this slow decay corresponds to long-range memory is still ongoing, econometricians tend to conclude that the statistical analysis in general cannot be expected to provide a definite answer concerning the presence or absence of long-range memory in asset price return \cite{Lo1991Econometrica,Willinger1999FinStoch,Mikosch2003}. 

Recently we have proposed an agent-based model, macroscopic dynamics of which can be reduced to Fokker-Planck equation or a set of stochastic differential equations (SDEs), which is able to precisely reproduce empirical probability density function (PDF) and power spectral density (PSD) of absolute return \cite{Kononovicius2013EPL,Gontis2014PlosOne} as well as scaling behavior of volatility return intervals \cite{Gontis2016PhysA}. Similar non-linearity can be introduced into GARCH(1,1) process \cite{Kononovicius2015PhysA} to recover $1/f$ noise as well. This raises a question if the long-range memory present in the financial markets should be embedded into the models or empirically observed property is just an outcome of non-linear agent interactions.

From our point of view there is a fundamental problem to empirically establish which of the possible alternatives, fBm or stochastic processes with non-stationary increments, is most well-suited to describe financial markets. The main idea is to employ the dependence of first passage time PDF on Hurst parameter $H$ for the fBm \cite{Ding1995PhysRevE,Metzler2014Springer}. This is in close connection with research of volatility return intervals \cite{Yamasaki2005PNAS,Wang2006PhysRevE,Wang2008PhysRevE,Bunde2011EPL,Bunde2014PRE,Gontis2008PhysA} and of first passage times in non-linear stochastic processes \cite{Gontis2012ACS}.

In Section \ref{sec:hurst-first-passage} we present the short theoretical background for our empirical investigation, in Section \ref{sec:empirical} we deal with empirical data from Forex and in Section \ref{sec:conclusion} we discuss and conclude results.

\section{Hurst parameter and first passage times of stochastic processes}
\label{sec:hurst-first-passage}

We will consider the first passage problem of stochastic processes $x(t)$ with absorbing boundary at some threshold level $x=h$. The problem for fBm was first considered by Ding and Yang as first return time \cite{Ding1995PhysRevE}.
 
\begin{figure}[h]
\centering
\includegraphics[width=\subfigf]{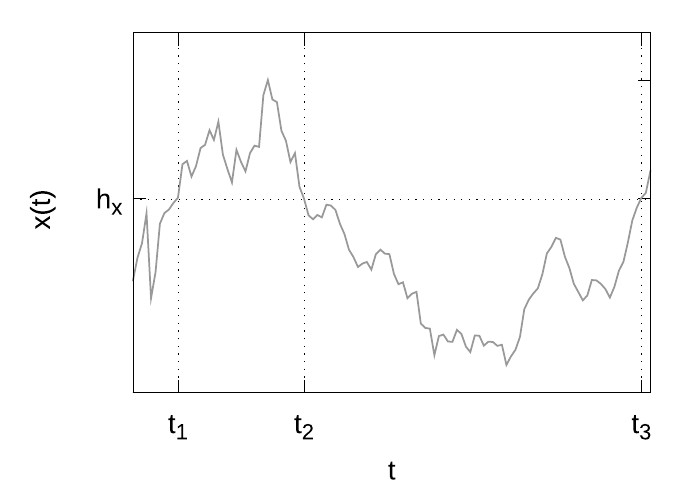}
\caption{\label{fig1} Exemplary fragment of a generic time series. Three threshold, $h_x$, passage events, $t_i$, are shown. Thus burst duration $T$ can be defined as $T=t_2-t_1$  and inter-burst duration $\theta$ can be defined as  $\theta=t_3-t_2$.}
\end{figure}

We consider two distinct threshold passage events -- one describes return to the threshold from above, while the other describes return to the threshold from below. Burst duration is the amount of time time series spent above the threshold (starts with passage from below and ends with passage from above; $T=t_2-t_1$  in \figref{fig1}). While the inter-burst duration is the amount of time series spent below the threshold (starts with passage from above and ends with passage from below; $\theta=t_3-t_2$ in \figref{fig1}). 
The PDFs of burst and inter-burst duration for fBm coincide and can be written as \cite{Ding1995PhysRevE,Metzler2014Springer} 
\begin{equation}
p(T) \sim T^{H-2}.
\label{eq:TPDF}
\end{equation}
Here $H$ is the Hurst parameter defining the exponent of power-law PDF $H-2$, which coincides with the corresponding exponent for other one-dimensional Markov processes only when $H=\frac{1}{2}$  \cite{Borodin2002Birkhauser,Jeanblanc2009Springer,Gardiner2009Springer,Redner2001Cambridge}. Burst and inter-burst duration can be defined through the first passage problem with initial value $x_0$ infinitesimally near the threshold $h$ \cite{Gontis2012ACS}.

It is widely accepted that the deviation of Hurst parameter from $H=\frac{1}{2}$ indicates about long-range memory in data, but this is not true as class of Markov processes with non-stationary increments can generate signals with $H\neq\frac{1}{2}$ and slowly decaying auto-correlation \cite{McCauley2006PhysA}.  We have earlier proposed to model financial markets by non-linear SDE, \cite{Gontis2004PhysA,Gontis2006JStatMech,Gontis2008PhysA,Gontis2014PlosOne,Gontis2016PhysA}, which belongs to this class as well. Many other agent-based models with non-linear interactions can lead to the macroscopic description by a non-linear SDEs of the following or a similar form\cite{Kaulakys2005PhysRevE,Ruseckas2014JStatMech} 
\begin{equation}
\rmd x=\left( \eta-\frac{\lambda}{2} \right) x^{2\eta-1} \rmd t +x^{\eta} \rmd W, \label{eq:SDE2}
\end{equation} 
having just two parameters:  $\eta$ as exponent of noise multiplicativity and $\lambda$ as exponent of power-law PDF. It was justified by various methods that this class of SDE generates the time series with the power-law behavior of stationary PDF and PSD \cite{Ruseckas2014JStatMech},
\begin{equation}
p(x)\sim x^{-\lambda},\qquad S(f)\sim \frac{1}{f^\beta},\quad \beta=1+\frac{\lambda-3}{2\eta-2}=2H+1.
\label{eq:power-law}
\end{equation}
It is noteworthy that these interconnected power-law properties are the direct outcome of scaling property in Fokker-Planck equation for two point transition probability $P(x\prime,t \vert x,t_0)$ corresponding to the SDE (\ref{eq:SDE2}) \cite{Ruseckas2014JStatMech}
\begin{equation}
aP(ax\prime,t \vert ax,0)=P(x\prime,a^{2(\eta-1)}t \vert x,0).
\label{eq:scaling}
\end{equation}
The class of SDEs (\ref{eq:SDE2}) describes multifractal stochastic processes with non-stationary increments \cite{McCauley2006PhysA,McCauley2007PhysA} and despite it's power-law auto-correlation and PSD (\ref{eq:power-law}) does not imply dynamics with correlated signals like those of fBm. Here we will look for the opportunity to establish test for the empirical data helping to make clear distinction between these two different models with correlated and  uncorrelated increments. 

The initial idea of such test comes from the PDF of first return times (\ref{eq:TPDF}) depending on $H$ in the case of fBm and corresponding PDF for non-linear SDEs (\ref{eq:SDE2}) characterized by power-law exponent $3/2$ common for all one-dimensional Markov processes \cite{Borodin2002Birkhauser,Jeanblanc2009Springer,Gardiner2009Springer,Redner2001Cambridge}. 
The burst duration for the non-linear SDEs (\ref{eq:SDE2}) were considered in \cite{Gontis2012ACS}. The asymptotic behavior of time $T$ PDF can be written in rather transparent form
\begin{gather}
p_{h_x}^{(\nu)}(T) \sim T^{-3/2}, \quad \text{for} \quad 0 < T \ll \frac{2}{(\eta-1)^2 h_x^{2(\eta-1)}j_{\nu,1}^2}, \\
p_{h_x}^{(\nu)}(T) \sim \frac{1}{T}\exp\left(-\frac{(\eta-1)^2 h_x^{2(\eta-1)}j_{\nu,1}^2 T}{2}\right) , \quad \text{for} \quad T \gg \frac{2}{(\eta-1)^2 h_x^{2(\eta-1)}j_{\nu,1}^2}.
\label{eq:burst}
\end{gather}
Here, $\nu=\frac{\lambda-2\nu+1}{2(\eta-1)}$, and $j_{\nu,1}$ is a first zero of a Bessel function of the first kind. The power-law behavior with exponent $3/2$ in Eq. (\ref{eq:burst}) is consistent with the general theory of the first passage times in one-dimensional stochastic processes \cite{Redner2001Cambridge,Jeanblanc2009Springer}. 

\section{Empirical investigation of burst and inter-burst duration}
\label{sec:empirical}

Empirical evaluation of PDF for burst and inter-burst duration should serve as a test for the nature of stochastic process. Deviations from the power-law with exponent $3/2$ in the case of fBm signals about true long-range dependence. Certainly, we will see later, such test is not so straightforward as real processes such as volatility in the financial markets may be more complicated than one-dimensional stochastic processes.

We investigate  burst and inter-burst duration for two financial variables: trading activity -- number of market trades during selected time window $\delta$ and volatility defined as filtered logarithmic return in the same time window. Following the idea of earlier proposed modeling, where trades arrive as in quasi-Poisson process driven by non-linear SDE \cite{Gontis2006JStatMech,Gontis2007PhysA,Gontis2008PhysA}, we will try to extract statistical properties of driving stochastic process from high frequency trading series in Forex. 

\begin{figure}[h]
\centering
\includegraphics[width=\subfigfff]{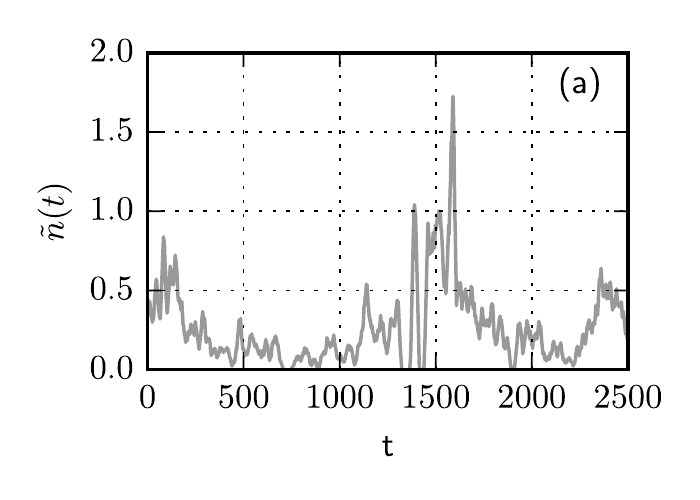}
\includegraphics[width=\subfigfff]{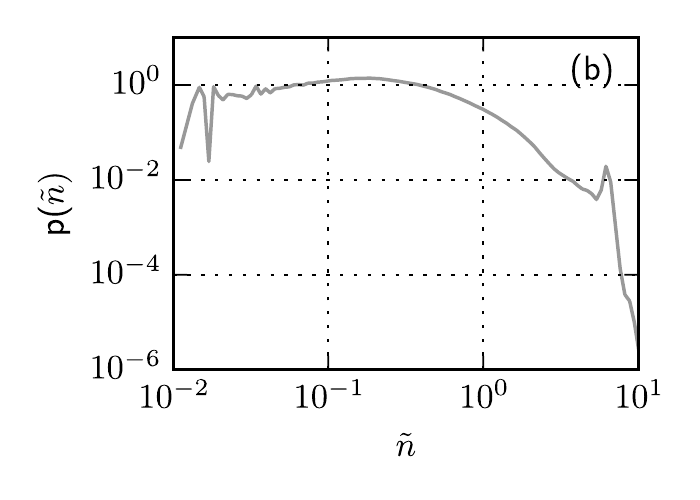}
\includegraphics[width=\subfigfff]{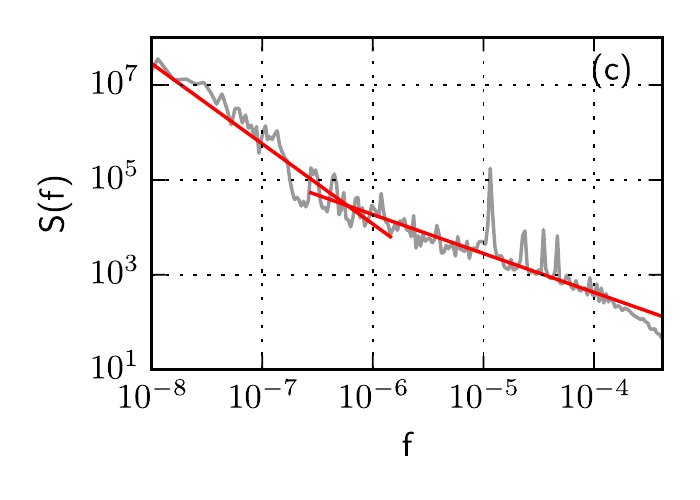}
\caption{\label{fig2} A fragment of EUR/USD filtered trading activity time series (a) as well as statistical properties, PDF (b) and PSD (c), of the whole time series recorded over the period of $10$ years. PSD (c) is approximated by two power-laws: one with exponent $\beta_1=1.7$ and the other with $\beta_2=0.8$.}
\end{figure}

First we investigate properties of trading activity recovered from the high frequency empirical data of EUR/USD exchange. These trading series have information about time of transactions executed. Time series are joint into the continuous time sequence recorded in the period of 10 years from January, 2000 to October, 2010, thus we are able to recover quasi-Poisson sequence of inter-trade duration's $\tau_p(i)$. According model assumption \cite{Gontis2007PhysA} these series are driven by continuous stochastic process $\tau(t)$ and the conditional probability of the inter-event time $\tau_p(i)$ is defined by 
the stochastic rate $\frac{1}{\tau}$ 
\begin{equation}
P(\tau_p(i)\vert \tau) = \frac{1}{\tau} \exp \left[ \frac{\tau_p(i)}{\tau} \right].
\label{eq:Poisson}
\end{equation}
One can define trading activity here as $n(t)=\frac{1}{\tau(t)}$, which exhibits long-range dependence, see \cite{Gontis2006JStatMech,Gontis2007PhysA,Gontis2008PhysA}. To recover $n(t)$ from empirical data we use Anscombe transform introduced to process image information \cite{Makitalo2013IEEE}, when signal intensity is low and Poisson fluctuations are considerable.  We evaluate the number of trades in the subsequent intervals of 60 seconds $n_{60}(t)$, apply procedure of moving average with time window $10$ after first Anscombe transform, then revert back with exact unbiased reverse Anscombe transform \cite{Makitalo2013IEEE}.  
The recovered stochastic empirical signal $\tilde{n}(t)$ is an object of our empirical investigation. It is easy to demonstrate that the PSD $S_n(f)$ of this signal exhibits long-range dependence with $\beta_1=1.7; \beta_2=0.8$, see \figref{fig2}(c).

The interpretation of this result is not trivial as $\beta_1 > 1$ and $\beta_2 < 1$. The recovered signal behaves as fBm and fractional Gaussian noise at the same time. Any additional statistical property of the signal might shed some light on the problem.  Pure fBm process should yield the PDF of  $T$ and $\theta$ according Eq. (\ref{eq:TPDF}). The value of Hurst parameter $H=(\beta_1-1)/2=0.35$ and the exponent of burst and inter-burst duration distribution $2-H=1.65$ should be expected if fBm was assumed as the model of long-range dependence. 

\begin{figure}[h]
\centering
\includegraphics[width=\subfigff]{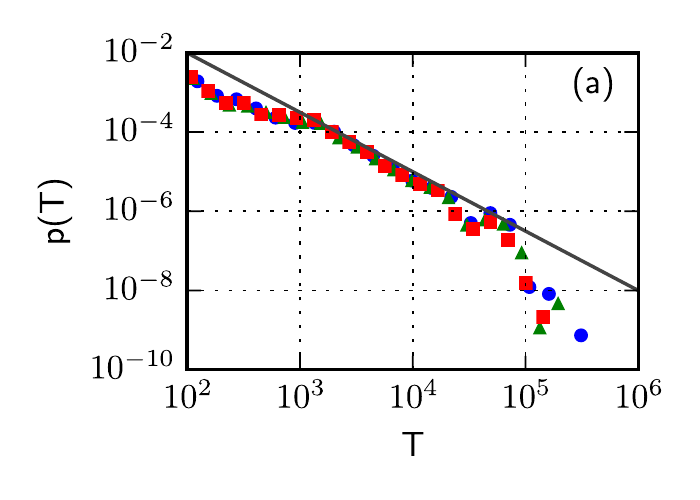}
\includegraphics[width=\subfigff]{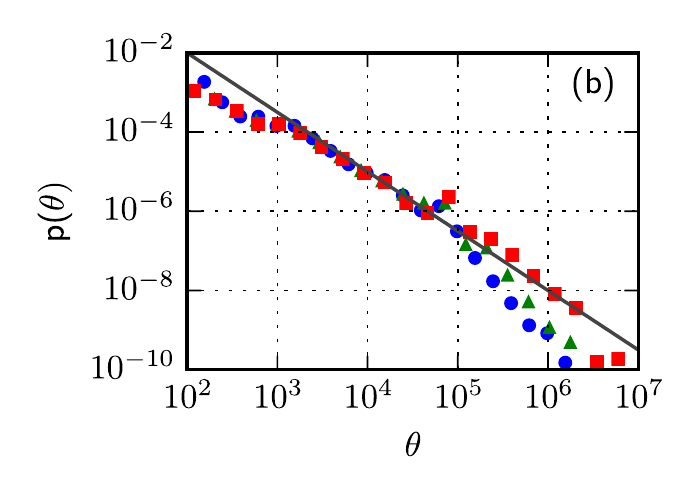}
\caption{\label{fig3} PDF of (a) burst and (b) inter-burst duration for EUR/USD filtered trading activity time series. Differently colored symbols (color online) represent different thresholds: (a) $h=0.3$ (blue circles), $0.4$ (green triangles), $0.67$ (red squares); (b) $h=1$ (blue circles), $1.5$ (green triangles), $2.5$ (red squares). In both cases black straight lines are power-law functions with exponent $3/2$.}
\end{figure}

In \figref{fig3} we demonstrate numerical results of $T$ and $\theta$ PDFs for empirical EUR/USD exchange trading activity $\tilde{n}(t)$.

The comprehensive interpretation of this result is a challenge, because empirical signal is very sophisticated. Even if we have succeeded to overcome the problem of the exogenous noise, Poisson fluctuations here, the signal probably incorporates at least double stochastic dynamics of agents and intraday seasonality, see \cite{Gontis2014PlosOne,Gontis2016PhysA}. Nevertheless,  from our point of view, the main long-range dependence in the financial markets, the part of PSD for the lowest frequencies, has to be recoverable from other more frequent fluctuations. Indeed, the PDF of burst duration $p(T)$, \figref{fig3}(a), exhibits power-law behavior in the region of two orders wide $600s\div6 60000s$ with exponent $3/2$. This is apparent only for low values of threshold  $h=0.3$, $0.4$, $0.67$, where bursts are wide and contribution of other noises is less considerable. For higher thresholds $h=1$, $1.5$, $2.5$, where inter-burst duration are bigger, the main power-law of $3/2$ appears valid for $p(\theta)$ up to three orders of magnitude of $\theta$ values, see \figref{fig3}(b). Thus we are able to recover the fundamental law of $3/2$ in the wide region of thresholds from 0.3 to 2.5. This law is a characteristic feature of one-dimensional Markov processes and observed result might signal that such process lies in the background of quasi-Poisson fluctuations for inter-trade duration Eq.(\ref{eq:Poisson}), \cite{Gontis2007PhysA}. Note that this long-range dynamics might have the same roots as dynamics of absolute return, see \cite{Gontis2014PlosOne} for more details.

Dealing with  volatility defined as filtered return we start from the very general modeling of return $r_{\delta}(t)$ time series by  
\begin{equation}
r_{\delta}(t) = \ln\frac{S(t+\delta)}{S(t)}=\sigma_t \omega_t,
\label{eq:return}
\end{equation}
where $S$ is an asset price, $\omega_t$ is a Gaussian noise with zero mean and unit variance, as in the family of ARCH  models of return series. In the agent-based modeling \cite{Gontis2014PlosOne,Gontis2016PhysA} Eq. (\ref{eq:return}) assumes interplay of exogenous noise $\omega_t$ with endogenous volatility $\sigma_t$ in very short time intervals $\delta$ of one minute order. When we are interested in the longer time scales $\Delta>>\delta$, $r_{\Delta}(t)=\Sigma r_{\delta}(t)$.
The volatility $\sigma_t$ in this modeling is  a linear function of the
absolute endogenous log price $\vert p(t) \vert$ derived as double stochastic Markov process from the Master equation of population dynamics for heterogeneous agents \cite{Gontis2014PlosOne}
\begin{equation}
\sigma_t = b_0(1+ a_0 \vert p(t) \vert).
\label{eq:defvolatil}
\end{equation}
Here $b_0$ normalizes standard deviation of $r_{\delta}(t)$ time series to the unit; $a_0$ is an empirical parameter which measures impact of endogenous dynamics on
the observed time series. The model, defined by Eqs. (\ref{eq:return})
and (\ref{eq:defvolatil}), comprises both the agent-based dynamic part described by
$\sigma_t$ and the exogenous noise part described by $\omega_t$. We will try to extract statistical properties of endogenous dynamics $\sigma_t$ from return time series in Forex. 

\begin{figure}[h]
\centering
\includegraphics[width=\subfigfff]{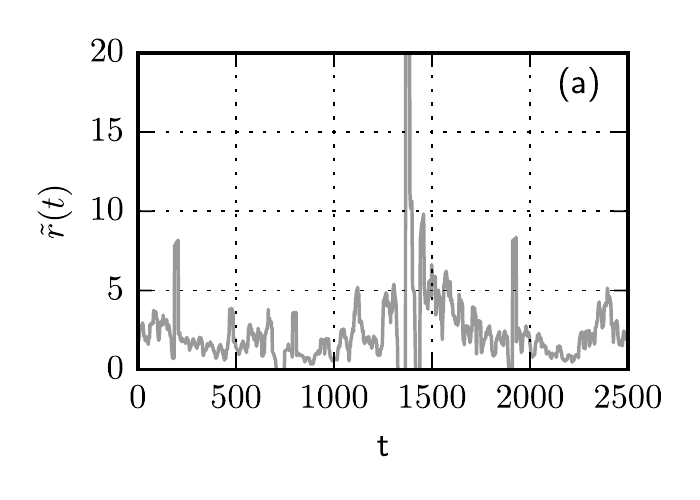}
\includegraphics[width=\subfigfff]{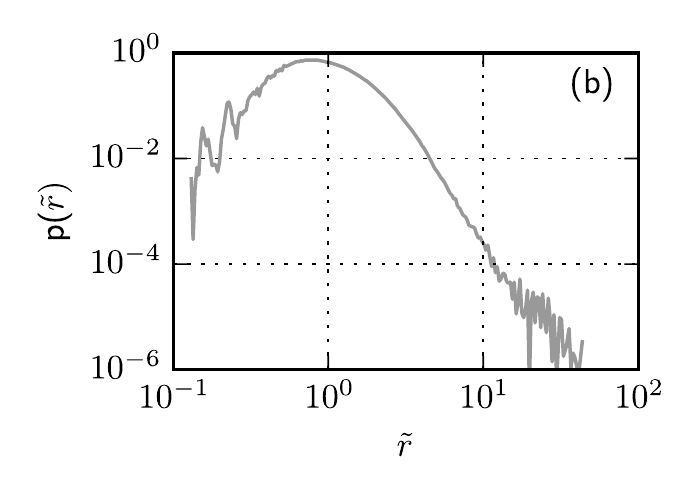}
\includegraphics[width=\subfigfff]{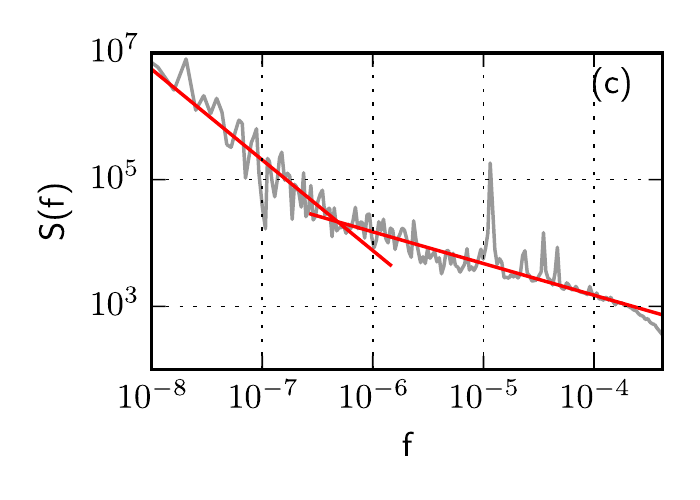}
\caption{\label{fig4} A fragment of EUR/USD filtered 1 minute return time series (a) as well as statistical properties, PDF (b) and PSD (c), of the whole time series recorded over the period of $10$ years. PSD (c) is approximated by two power-laws: one with exponent $\beta_1=1.4$ and the other with $\beta_2=0.5$.}
\end{figure}

Let us now consider empirical time series of EUR/USD return $r_{\delta}(t)$ with $\delta=60 s$. Having in mind quite general assumption about exogenous noise $\omega$ Eq. (\ref{eq:return}), we filter it by standard deviation filter with time window of 10 min. Statistical properties of these time series are given in \figref{fig4}.

The main statistical properties for return series are slightly different from trading activity, for example, values of exponents for PSD are  $\beta_1=1.4$ and $\beta_2=0.5$. The value of Hurst parameter $H=(\beta_1-1)/2=0.2$ thus the exponent of corresponding burst and inter-burst duration distribution $2-H=1.8$ should be expected. This implies a meaningful deviation from $3/2$ law, which has to be observable in the PDFs of burst and inter-burst duration. 

\begin{figure}[h]
\centering
\includegraphics[width=\subfigff]{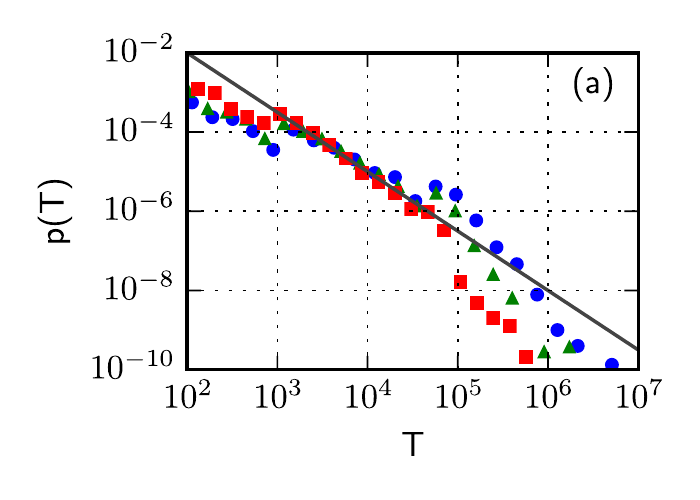}
\includegraphics[width=\subfigff]{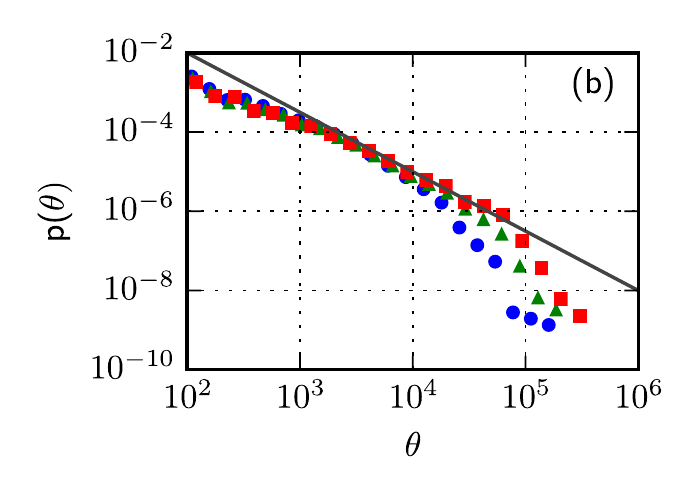}
\caption{\label{fig5} PDF of (a) burst and (b) inter-burst duration for EUR/USD filtered 1 minute return time series. Differently colored symbols (color online) represent different thresholds: (a) $h=0.3$ (blue circles), $0.4$ (green triangles), $0.67$ (red squares); (b) $h=1$ (blue circles), $1.5$ (green triangles), $2$ (red squares). In both cases black straight lines are power-law functions with exponent $3/2$.}
\end{figure}

In \figref{fig5}(a) the PDF of burst duration $p(T)$, exhibits power-law $3/2$ behavior in the region of two orders wide for low values of threshold  $h=0.3$, $0.4$, $0.67$. The cut of of power-law starts earlier for higher values of the threshold. For higher thresholds $h=1$, $1.5$, $2$, see \figref{fig5}(b), where inter-burst duration are bigger, the main power-law of $3/2$ appears valid without any intention to approach exponent $1.8$, which would be more appropriate for fBm process. Thus we are able to recover the fundamental law of $3/2$ in the wide region of thresholds for return series as well. This implies that supposed long-range dependence of return series probably is driven by one-dimensional Markov process, derived in the model of finance with heterogeneous agents  \cite{Gontis2014PlosOne,Gontis2016PhysA}.

\begin{figure}[h]
\centering
\includegraphics[width=\subfigfff]{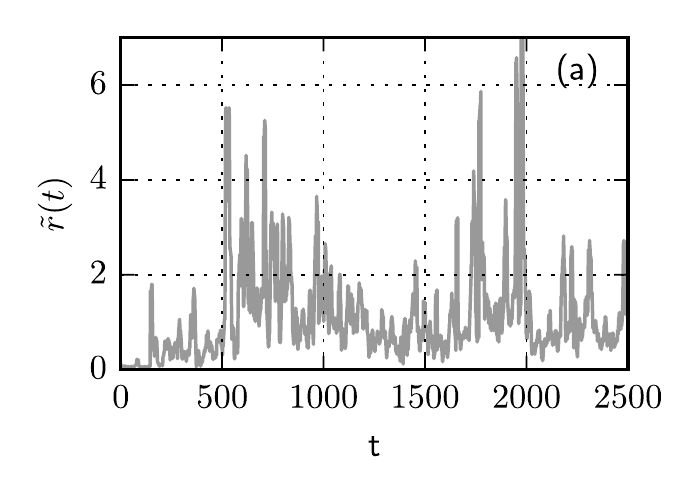}
\includegraphics[width=\subfigfff]{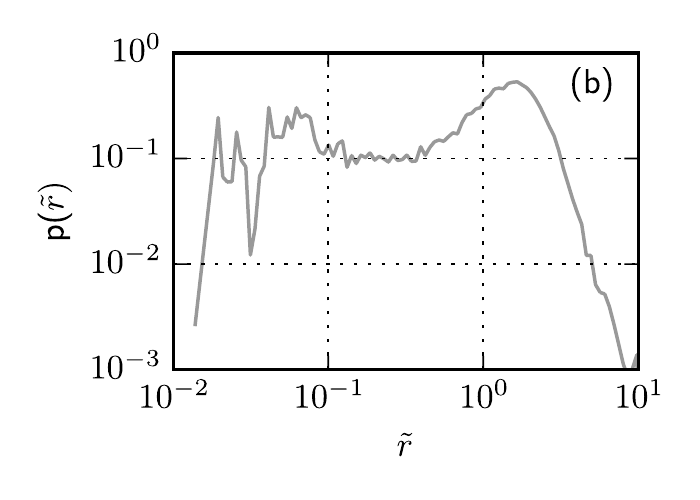}
\includegraphics[width=\subfigfff]{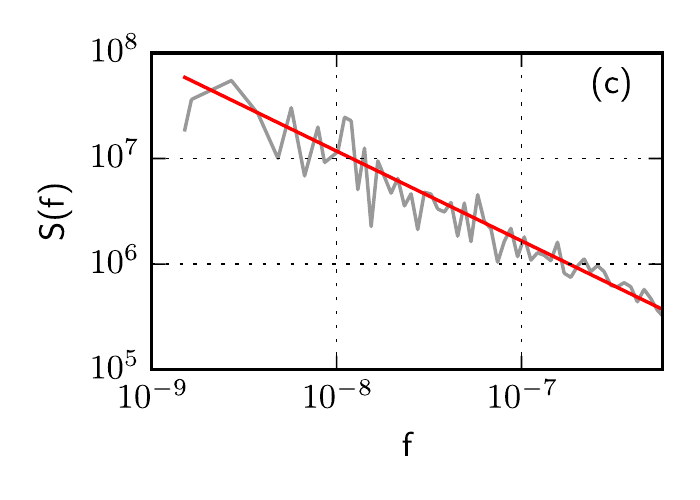}
\caption{\label{fig6} A fragment of NOK/USD filtered daily return time series (a) as well as average statistical properties, PDF (b) and PSD (c), of the whole time series of $5$ exchange rates recorded over the period of $50$ years. PSD (subfigure (c)) is approximated by a single power-law with exponent $\beta=0.9$.}
\end{figure}

The procedures of signal filtering in two previous examples of financial time series helped us to diminish the influence of exogenous noise $\omega_t$. In the case of longer time scales of return definition $\Delta\gg\delta$, when $r_{\Delta}=\Sigma r_{\delta}$, the exogenous noise becomes integrated into $r_{\Delta}(t)$. It is an open question how the exogenous noise impacts PDFs of burst and inter-burst duration for the $\tilde{r}_{\Delta}(t)$ calculated with standard deviation  filter applied to the empirical series $r_{\Delta}(t)$. We analyze the daily return series for CHF/USD, DKK/USD, JPY/USD, NOK/USD, USD/GBP exchanges in the period of $50$ years and apply standard deviation filter with time window $10$ trading days. In \figref{fig6} we present the main statistical properties of NOK/USD daily return series.

\begin{figure}[h]
\centering
\includegraphics[width=\subfigff]{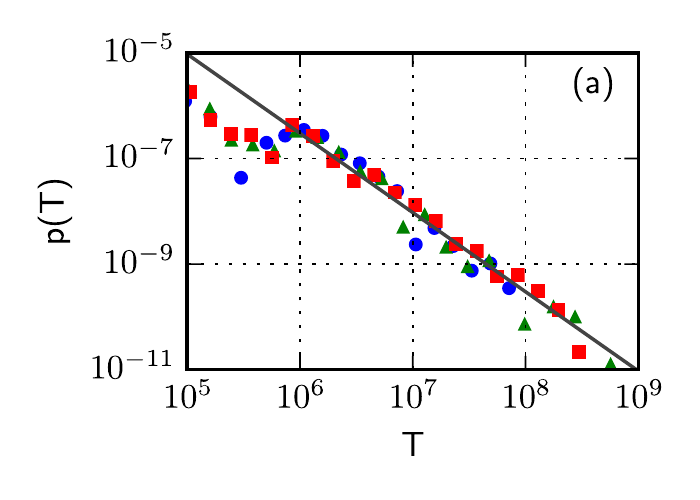}
\includegraphics[width=\subfigff]{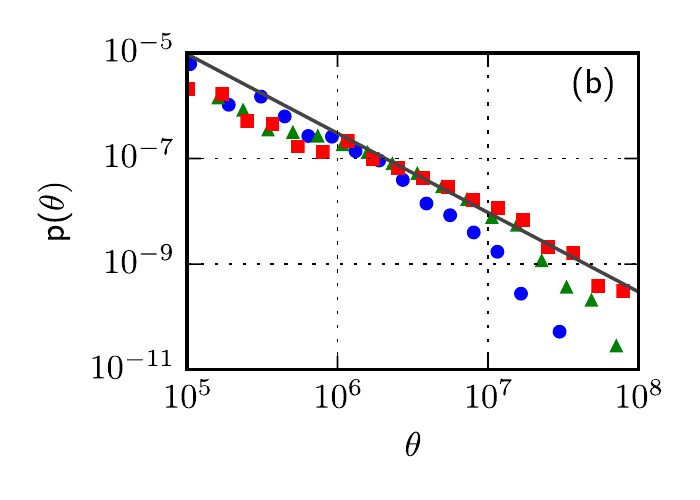}
\caption{\label{fig7} PDF of (a) burst and (b) inter-burst duration for the filtered daily return time series of $5$ exchange rates. Differently colored symbols (color online) represent different thresholds: (a) $h=0.3$ (blue circles), $0.4$ (green triangles), $0.67$ (red squares); (b) $h=1.5$ (blue circles), $2.5$ (green triangles), $3$ (red squares). In both cases black straight lines are power-law functions with exponent $3/2$.}
\end{figure}

The main peculiarity of daily series is the PSD with only one power-law exponent $\beta=0.9$, see \figref{fig6}(c). The value of $\beta$ lower than 1 is a consequence of exogenous and other frequent fluctuations integrated into the daily return. However, PDFs of burst and inter-burst duration defined for these series can be easily calculated as well. We demonstrate these PDFs calculated as $T$ and $\theta$ histograms in \figref{fig7}. Though the amount of data is limited and histograms are noise, PDFs are consistent with $3/2$ power-law. 

\section{Discussion and Conclusions}
\label{sec:conclusion}

The fluctuations of volatility and trading activity in the financial markets are known to exhibit $1/f$ noise or slowly decaying auto-correlation \cite{Engle2001QF,Plerou2001QF,Gabaix2003Nature,Ding2003Springer}. It becomes evident that the explanation of this statistical property is related with the model assumption of the financial system. The most frequent exogenous fluctuations can be filtered from empirical data in order to reveal  peculiarities of expected long-range behavior \cite{Gontis2016APPA}. Time series of volatility and trading activity, from our point of view, is a result of heterogeneous agent dynamics, which in macroscopic level can be described by Fokker-Plank equation or a set of SDEs \cite{Kononovicius2013EPL,Gontis2014PlosOne,Gontis2016PhysA}. Ordinary SDEs driven by Wienner noise describe Markov processes and usually are not treated as suitable to model long-range memory. Stochastic processes driven by fBm could be considered as possible alternative. Thus the choice between these two possibilities is the fundamental question for understanding and modeling of the financial markets.

Here we proposed a possible test for these alternatives investigating burst and inter-burst duration of empirical time series. For fBm process it is well known that PDF of first passage time is dependent on $H$, $p(T) \sim T^{H-2}$ \cite{Ding1995PhysRevE,Metzler2014Springer}. On the other hand for Markov processes it is well known that first passage time should scale as $p(T) \sim T^{-3/2}$ \cite{Redner2001Cambridge}. Though the dynamics of heterogeneous agents might result in multivariate statistics, when we need more than two groups of different agents, sometimes it is possible to segregate stochastic processes by time scale, when frequencies of fluctuations in various agent groups are considerably different. From our point of view this is the case in the financial markets and PSD with few values of $\beta$ can be modeled with such assumption \cite{Gontis2014PlosOne,Gontis2016PhysA}.

Instead of details in agent-based modeling, here we just wanted to establish whether macroscopic modeling by non-linear SDEs can be reasonable. The main result of empirical investigation of PDF for burst and inter-burst duration suggests that power-law part is consistent with the exponent $3/2$, which is characteristic feature of one-dimensional stochastic processes generated by ordinary SDE including non-linear \cite{Redner2001Cambridge}.

The result is confirmed for $1$ minute EUR/USD return and trading activity series and for CHF/USD, DKK/USD, JPY/USD, NOK/USD, USD/GBP daily return series. This is in agreement with previous research of volatility return intervals \cite{Gontis2016PhysA}.

Though there is no independent agent-based model for trading activity, our result confirms that such models should be consistent with description by non-linear SDEs as well. Also it confirms that PDFs of burst and inter-burst duration are in agreement with  power-law relation of  trading activity and absolute return assumed in the modeling of the financial markets \cite{Gontis2014PlosOne,Gontis2016PhysA}.

The most general conclusion of this empirical investigation is an opportunity to explain the so-called long-range memory in the financial markets by ordinary non-linear SDEs, Eq. (\ref{eq:SDE2}), representing multifractal stochastic processes with non-stationary increments. This is a very realistic alternative to the modeling incorporating fBm. Further empirical analyses of burst and inter-burst duration for various markets and assets should considerably strengthen and extend these conclusions.

\end{document}